\documentstyle[multicol,eqsecnum,aps,epsf]{revtex}
\newcommand{\be}{\begin{equation}}
\newcommand{\ee}{\end{equation}}
\newcommand{\bea}{\begin{eqnarray}}
\newcommand{\eea}{\end{eqnarray}}
\begin{document}
\draft
\title{The non-centrosymmetric lamellar phase in blends of \\
ABC  triblock and ac
diblock copolymers}
\author{Robert A. Wickham\cite{BYLINE1} and An-Chang Shi \cite{BYLINE2}}
\address{Department of Physics and Astronomy, McMaster University, Hamilton ON
L8S 4M1, Canada}
\date{\today}
\maketitle
%
%
\begin{abstract}

The  phase  behaviour  of  blends  of  ABC  triblock  and  ac  diblock
copolymers  is examined using  self-consistent field  theory.  Several
equilibrium lamellar structures are  observed, depending on the volume
fraction of the  diblocks, $\bar{\phi}_{2}$, the monomer interactions,
and the degrees of polymerization of the copolymers.  For segregations
just above  the order-disorder  transition the triblocks  and diblocks
mix together to form  centrosymmetric lamellae.  As the segregation is
increased  the triblocks  and  diblocks spatially  separate either  by
macrophase-separating, or by forming a non-centrosymmetric (NCS) phase
of alternating layers of  triblock and diblock (...ABCcaABCca...). The
NCS    phase    is    stable     over    a    narrow    region    near
$\bar{\phi}_{2}=0.4$. This region is widest near the critical point on
the phase coexistence curve and narrows to terminate at a triple point
at  higher segregation.  Above  the triple  point  there is  two-phase
coexistence  between almost  pure  triblock and  diblock phases.   The
theoretical phase diagram is consistent with experiments.

\end{abstract}
\begin{multicols}{2}
%
%
\section{Introduction}

Materials  which  lack  a centre  of  symmetry  in  the absence  of  a
polarizing field  are rare in  nature, and have attracted  much recent
interest   \cite{PETSCHEK87,TOURNILHAC92,STUPP97,GOLDACKER99}.   These
non-centrosymmetric (NCS)  materials can exhibit  dipolar second-order
nonlinear    optical    activity    (second-harmonic    generation)
\cite{TOURNILHAC92,STUPP97},  in   addition  to  piezoelectricity  and
pyroelectricity  \cite{TOURNILHAC92},  without  the  need to  apply  a
polarizing field.  As such, NCS  materials are of  great technological
interest.  Recently, the capability  to make  NCS structures  in block
copolymer  blends  has  been  demonstrated  experimentally 
\cite{GOLDACKER99}.  Block
copolymers consist of  two (or more) chains, or  blocks, of chemically
distinct  monomers  covalently  bonded  end-to-end to  form  a  single
polymer. Competition between the repulsion of unlike blocks and the 
constraint that the blocks are attached together leads to the formation
of ordered periodic structures.
Block copolymers are promising materials to use in the design
of NCS structures  due to the high degree of control  one has over the
structure and properties of  the blocks.  NCS structures created using
polymers have longer periods than those created previously using small
molecules   \cite{TOURNILHAC92,STUPP97}.   The   periodicity  of   the
structure can  be changed  by adjusting the  block size,  creating the
potential to tune the  wavelengths for second-harmonic generation. The
dielectric properties  of the  blocks can be  tailored to  the desired
application.     Finally, block copolymers self-assemble into
periodic NCS  structures, so no microscale  fabrication techniques are
necessary to produce the NCS structure.

The key breakthrough in Ref.\ \cite{GOLDACKER99} that made possible the
formation of  NCS structures in  block copolymers was  the recognition
that  blends of  ABC triblock  copolymers and  ac  diblock copolymers,
instead  of  pure  melts  of  ABC  triblock  copolymers  are  required
\cite{EXCEPTION1}.  Here A,  B and C refer to  the chemical species of
each block --- the a and c blocks on the diblock are the  same chemical
species as the  A and C blocks on the triblock.  Pure ac diblock melts
produce stable lamellar,  hexagonally-packed cylindrical, body-centred
cubic,  and   gyroid  phases  ---   all  of  which  have   centres  of
symmetry. Pure  ABC triblock melts  have an even richer  phase diagram
(see Refs.\ \cite{ZHENG95} and  \cite{ABETZ00} for a  discussion), but
almost all  of the  phases so  far  discovered are centrosymmetic  (CS)
\cite{EXCEPTION2}. Compared to the pure phases, the behaviour of blends 
of triblocks and diblocks is relatively unexplored and is a topic of 
current fundamental interest.

In Ref.\ \cite{GOLDACKER99} only
lamellar structures were reported, and we will restrict ourselves to
discussing  lamellar  structures  in   this  paper.  As  discussed  in
Refs.\ \cite{GOLDACKER99}  and  \cite{LEIBLER99}  and  shown  here  in
Fig.\ 1,  possible lamellar structures  in blends of
ABC  triblock and  ac  diblock copolymers  include: triblock-rich  and
diblock-rich  phases where the  triblock and  diblock mix  together to
form a  ``mixed'' centrosymmetric structure (MCS phase),  a NCS phase
where the  triblocks and diblocks spatially  separate into alternating
triblock  and  diblock  layers (...ABCcaABCca...),  a  centrosymmetric
double-layer phase (CS phase) of alternating double-layers of triblock
and diblock  (...ABCcaacCBA...), and regions  of two-phase coexistence
between these phases. When it is favourable for the triblocks and diblocks 
to spatially  separate, it is unclear whether the system will achieve 
this by forming a structure with alternating triblock  and  
diblock  layers, or by macrophase separation. However, by carefully tuning 
the system parameters the authors of Ref.\ \cite{GOLDACKER99} were able to 
find the NCS  structure.

Given the complexity  of the pure ABC triblock  phase diagram, and the
even  greater  complexity  of  the  phase diagram  for  the  blend,  a
theoretical guide to experimental searches for the NCS structure would
be helpful. The two main questions one would like to answer are: ``What
is the driving mechanism behind the formation of the NCS phase?'' and
``Where in phase space should one expect the NCS structure to be stable?''
To date, the only theoretical studies of these blends are those    of   
Leibler   {\em    et   al.}   \cite{LEIBLER99}  and Birshtein {\em et al.}
\cite{BIRSHTEIN99} in   the
strong-segregation limit. These papers focused on answering the
first question by explaining the stability  of the NCS phase in terms
of an entropic advantage to forming mixed aA domains of a and A block 
(from the diblock and the triblock, respectively), when compared with the 
formation of AA (and aa) domains. However, these studies do not directly 
address the second question.
 In this paper, we  expand 
the scope of theoretical understanding 
by examining blends of ABC   triblock  and  ac   
diblock  copolymers   using  self-consistent
field  theory. Although this formalism can be used to determine in detail the
structure of the aA interfaces \cite{SHI94}, 
thereby providing information about 
the driving mechanism, we do not focus on this here. Rather,
our aim is to  answer the second question 
by  examining the effect of blend composition on the phase behaviour, the 
possibility for phase-separation, and the phase behaviour  in the weak  
to intermediate segregation regime. 
%
%
\begin{figure}
\begin{center}
    \leavevmode
    \epsfxsize=2.5 in
    \epsfysize=5 in
    \epsffile{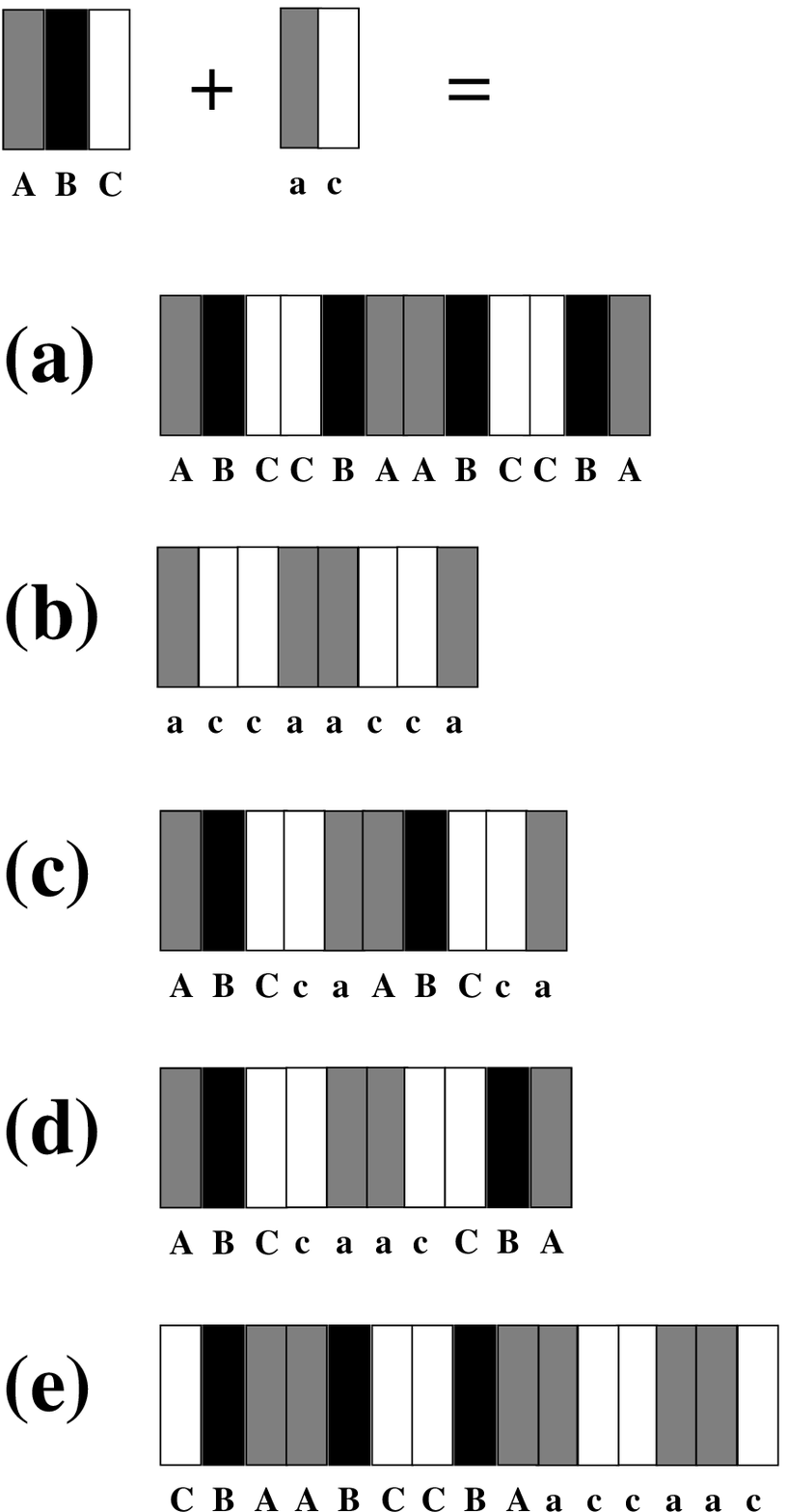}
\end{center}
{\small FIG.\ 1.
Some possible lamellar structures in a blend of ABC triblock and ac
diblock copolymers: (a) triblock-rich mixed centrosymmetric, 
(b) diblock-rich mixed centrosymmetric,
(c) non-centrosymmetric, (d) centrosymmetric 
double-layer phase, (e) two-phase coexistence.}
\label{FIG:STRUCTURES}
\end{figure}
%
%
%
\section{Mean-Field Theory}

Consider  an incompressible  blend of  ac diblock  copolymers  and ABC
triblock  copolymers in a volume $V$.   
The  total  degree of  polymerization  of  the
diblock  is $N$; for  the triblock  it is  $\Omega N$ (thus $\Omega$ is 
the ratio of the triblock degree of  polymerization to that of the 
diblock). The diblock
consists  of  an a-block,  with  a  degree  of polymerization  $f_{2A}
N$,  and  a  c-block,  with  a degree  of  polymerization  $f_{2C}
N$. The  diblock composition variables satisfy $f_{2A}  + f_{2C} =
1$.  Similarly, the triblock consists  of three blocks with degrees of
polymerization $f_{3 \alpha}  \Omega N$, where $\alpha =$ A,  B or C. The
triblock  composition variables satisfy  $f_{3A} +  f_{3B} +  f_{3C} =
1$.  We scale  distances by  the Gaussian  radius of  gyration  of the
diblock, $R_{g2}  = b  ( N /  6)^{1/2}$.  The  monomer statistical
Kuhn length $b$ and the bulk monomer density $\rho_{0}$ are assumed to
be the same for all three  chemical species.  In what follows, we will
scale  the chain arc-length  by the  diblock degree  of polymerization
$N$.

Beginning with  the many-chain Edwards Hamiltonian, we  can derive the
free-energy  $F$   of  the  blend  in   the  mean-field  approximation
\cite{HELFAND75}.  The suitably-scaled  free-energy density $f$ at temperature
$T$ has the form
%
%
\bea
f \equiv \frac{N F}{\rho_{0} V k_{B} T} & = & \frac{1}{V} 
\int d {\bf r} \Big{\{}
\chi_{AB} N \mbox{ } \phi_{A} ({\bf r}) \phi_{B}({\bf r}) 
\nonumber  \\ & &  \hspace{-.75 in} 
+ \chi_{AC} N \mbox{ } \phi_{A} ({\bf r}) \phi_{C}({\bf r}) + 
\chi_{BC} N \mbox{ } \phi_{B} ({\bf r}) \phi_{C}({\bf r}) 
\nonumber  \\ & &  \hspace{-1 in}
- \sum_{\alpha = A,B,C} \omega_{\alpha} ({\bf r}) \phi_{\alpha} 
({\bf r}) \Big{\}}
 - e^{\mu_{2}} \mbox{ } Q_{2}[\omega] 
- Q_{3}[\omega]. 
\label{EQ:FREE}
\eea
We derived  Eqn.\ (\ref{EQ:FREE}) using the grand canonical ensemble 
\cite{MATSEN95}. Multiple phase-coexistence, which we will encounter below,
is most conveniently studied using this ensemble. The 
three Flory-Huggins interaction parameters  $ \chi_{AB}$, $\chi_{AC}$, 
and $\chi_{BC}$ are responsible for repulsion between unlike blocks.
We denote the volume fraction of $\alpha$ monomers from the $n$-blocks at 
position ${\bf r}$ as
$\phi_{n \alpha} ({\bf r})$ and write the volume fraction of 
$\alpha$ monomers, $\phi_{\alpha} ({\bf r})$, and of diblocks, 
$\phi_{2}({\bf r})$, as
%
%
\bea
\phi_{\alpha} ({\bf r}) & \equiv & \phi_{2 \alpha} ({\bf r})
			 + \phi_{3 \alpha} ({\bf r}) \\
\phi_{2}({\bf r})  & \equiv & \phi_{2A} ({\bf r}) + \phi_{2C} ({\bf r}).
\eea
The chemical potential for the diblocks is 
$\mu_{2}$, in units of $k_{B} T$. Since the blend is 
incompressible, the chemical potential for the triblocks
can be set to zero without loss of generality.

In Eqn.\ (\ref{EQ:FREE}), $Q_{2}[\omega]$ is the partition function of 
a single diblock copolymer interacting with the mean-fields 
$\omega_{\alpha}({\bf r})$.
Similarly, $Q_{3}[\omega]$ is the partition function of a single 
triblock copolymer interacting with these mean-fields. These partition 
functions may be written in  terms of the propagators 
$Q_{\alpha} ({\bf r}, s |  {\bf r}')$, which give the 
probability that the $\alpha$ monomer at
arc-length $s$ is at
position ${\bf r}$, given that the $\alpha$ monomer 
at arc-length 0 is at ${\bf r}'$:
%
%
\bea
Q_{2}[\omega] & = & \frac{1}{V} \int d {\bf r}_{1} d {\bf r}_{2} d {\bf r}_{3}
\mbox{ } Q_{A} ({\bf r}_{1}, f_{2A} | {\bf r}_{2}) 
\mbox{ } Q_{C} ({\bf r}_{2}, f_{2C} | {\bf r}_{3}) 
\label{EQ:Q2} 
\\
Q_{3}[\omega] & = & \frac{1}{V} \int d {\bf r}_{1} d {\bf r}_{2} d {\bf r}_{3}
d {\bf r}_{4}
\mbox{ } Q_{A} ({\bf r}_{1}, f_{3A} \Omega | {\bf r}_{2}) 
\nonumber \\ & & \hspace{.25 in} 
\times
\mbox{ } Q_{B} ({\bf r}_{2}, f_{3B} \Omega | {\bf r}_{3}) 
\mbox{ } Q_{C} ({\bf r}_{3}, f_{3C} \Omega | {\bf r}_{4}). 
\eea
The factors of $1/V$ are inserted above for convenience.
The propagators satisfy the modified diffusion equation
%
%
\be
\frac{\partial}{\partial s} Q_{\alpha} ({\bf r}, s |{\bf r'}) = 
\nabla^{2}_{{\bf r}}  Q_{\alpha} ({\bf r}, s |{\bf r'}) - \omega_{\alpha} 
({\bf r})
Q_{\alpha} ({\bf r}, s |{\bf r'})
\label{EQ:DIFFUSION}
\ee
with the initial condition
%
%
\be
Q_{\alpha} ({\bf r}, 0|{\bf r'}) = \delta ( {\bf r} - {\bf r'}).
\label{EQ:INITIAL}
\ee

In the mean-field approximation the fields $\omega_{\alpha}$
are related to the monomer volume fractions through the relations 
%
%
\bea
\label{EQ:MFB}
\omega_{A} ({\bf r}) & = & \chi_{AB} N \mbox{ } \phi_{B} ({\bf r}) 
			+ \chi_{AC} N \mbox{ } \phi_{C} ({\bf r}) 
			+ \eta ({\bf r}) \\
\omega_{B} ({\bf r}) & = & \chi_{AB} N \mbox{ } \phi_{A} ({\bf r}) 
			+ \chi_{BC} N \mbox{ } \phi_{C} ({\bf r}) 
			+ \eta ({\bf r}) \\
\omega_{C} ({\bf r}) & = & \chi_{AC} N \mbox{ } \phi_{A} ({\bf r}) 
			+ \chi_{BC} N \mbox{ } \phi_{B} ({\bf r}) 
			+ \eta ({\bf r}) ,
\eea
where the field $\eta$ is to be adjusted to 
enforce the incompressibility condition
%
%
\be
\phi_{A} ({\bf r}) + \phi_{B} ({\bf r}) + \phi_{C} ({\bf r}) = 1. 
\ee
The monomer volume fractions are, in turn, related to
functional derivatives of  $Q_{2}[\omega]$ and $Q_{3}[\omega]$:
\bea
\phi_{2\alpha} ({\bf r}) & = & - V \mbox{ } e^{\mu_{2}} \mbox{ }
\frac{ \delta Q_{2} [\omega] }{
\delta \omega_{\alpha} ({\bf r}) }  
\\
\phi_{3\alpha} ({\bf r}) & = &
- V \mbox{ } 
\frac{ \delta Q_{3} [\omega] }{\delta \omega_{\alpha} ({\bf r}) } .
\label{EQ:MFE}
\eea
These functional derivatives are evaluated using 
Eqns.\ (\ref{EQ:Q2}--\ref{EQ:INITIAL}).

To obtain  the {\em exact} mean-field  solution for a  given point in
parameter space, Eqns.\   (\ref{EQ:MFB}--\ref{EQ:MFE}) need to be solved
self-consistently  using numerical  methods.  The  method  of solution
involves selecting a  set of basis functions appropriate  to the space
group of the periodic structure  to be examined, and reformulating the
theory   in   the   reciprocal   space  of   these   basis   functions
\cite{MATSEN94}.  With  initial guesses for the  periodicity, $D$, and
monomer  profiles, $\phi_{\alpha}$, of  the structure,  the reciprocal
space  versions  of   Eqns.\  (\ref{EQ:MFB}--\ref{EQ:MFE})  are  solved
iteratively  to obtain  the  mean-field profiles  and free-energy 
density  $f$
corresponding  to the  chosen  $D$.  This  procedure  is repeated  for
different choices  of $D$  until the free-energy density 
 is minimized  at the
system's  preferred  periodicity.  The  numerical  procedure  and  the
reciprocal space  formulation are discussed  in more detail  in Refs.\
\cite{MATSEN94,SHI96,LARADJI97}.   The preferred  periodicity  and
minimal free-energy 
density are determined for all possible structures.  These
free-energy densities 
 are compared and the  structure with the lowest $f$ at a
given  point in  phase  space  is the  equilibrium  structure at  that
point.  In   the  $f$--$\mu_{2}$  plane,   two-phase  coexistence  
occurs when the free-energy  density 
curves  for  two
structures  cross at a  given  value of  $\mu_{2}$. Three-phase  coexistence
occurs when  three such curves intersect  at a point.   Even though we
compare free-energy density  curves as a  function of $\mu_{2}$, when  we plot
phase   diagrams  we   use   the  average   diblock  volume   fraction,
$\bar{\phi}_{2}$,   as  a  variable,   instead  of   its  thermodynamic
conjugate, $\mu_{2}$ \cite{PHI2}.

A  pure  triblock copolymer  melt  can  form  many different  periodic
structures \cite{ZHENG95,ABETZ00}.  The number of structures formed by
blending  triblock copolymers  with  diblock copolymers  is even  more
diverse. In  this paper, we restrict ourselves  to discussing lamellar
structures,  since   they  have  been  the   subject  of  experimental
investigations of  non-centrosymmetry \cite{GOLDACKER99}.  The problem
is then one-dimensional with cosines and sines as basis functions. The
non-centrosymmetric phase  was obtained  using both sines  and cosines
and an  initial monomer profile  that was NCS (...ABCca...)   to begin
the iteration procedure.  Centrosymmetric  phases have only cosines as
basis functions.  The  MCS structure was found at  about the period of
the NCS  structure. A centrosymmetric  double-layer structure was
found   at  about   twice   the  NCS   period,   using  the   sequence
(...ABCcaacCBA...) as  an initial profile for the  iteration step.  
The number of basis functions in our computation of the free-energy density 
is selected to achieve   an accuracy of  $10^{-6}$, which  is more
than  sufficient  to  resolve  the small  differences  in  free-energy
between these phases.  As the blend segregation increases
 the interfaces
become  sharper and  more  basis  functions are  needed  to achieve  this
accuracy.

%
%
\section{Results and Discussion}

Since  the parameter  space for  this system  is large  -- $\chi_{AB}
N$,  $\chi_{AC}  N$,  $\chi_{BC} N$,  $\Omega$,  $f_{2A}$,
$f_{3A}$,   $f_{3B}$   and   $\bar{\phi}_{2}$   can  all   be   varied
independently --  we have  to be careful  to select  parameters which
favour  a stable  lamellar phase.   A  stable lamellar  phase is  most
likely  when  the  block  compositions  are  symmetric,  $f_{2A}=1/2$,
$f_{3A}=f_{3B}=1/3$, and the  Flory-Huggins interaction parameters are
equal,  $\chi_{AB}=\chi_{AC}=\chi_{BC}\equiv  \chi$.  Accordingly,  we
fix  the  block  compositions   to  be  symmetric,  and  examine  only
situations where the Flory-Huggins interaction parameters are equal or
nearly equal.  Even within  these bounds, the  phase behaviour  of the
blend is rich.

The phase diagram  in the $\chi N$---$\bar{\phi}_{2}$ plane, for
the case where all the Flory-Huggins interaction parameters are equal,
is shown in  Fig.\ 2. The ratio of  triblock to diblock
lengths  is $\Omega=1.5$.   At the lowest values of $\chi N$  the  blend is
disordered.  As $\chi N$ increases there is a  transition to a
mixed centrosymmetric (MCS) phase where the triblocks and diblocks mix
together uniformly, but form  a lamellar phase \cite{BIRSHTEIN99(2)}. 
If $\bar{\phi}_{2}$ is
small, and the blend is  triblock-rich, the lamellar structure will be
alternating    triblock   layers    (...ABCCBA...),   as    in   Fig.\
1a.    As  $\bar{\phi}_{2}$  increases   the  blend
becomes  diblock-rich and  the structure  becomes  alternating diblock
layers  (...acca...), as  in Fig.\  1b.   A typical
density profile for the MCS phase is shown in Fig.\ 3.
\vspace{.3 in}
%
%
\begin{figure}
\begin{center}
    \leavevmode
    \epsfxsize=3.0 in
    \epsfysize=3.0 in
    \epsffile{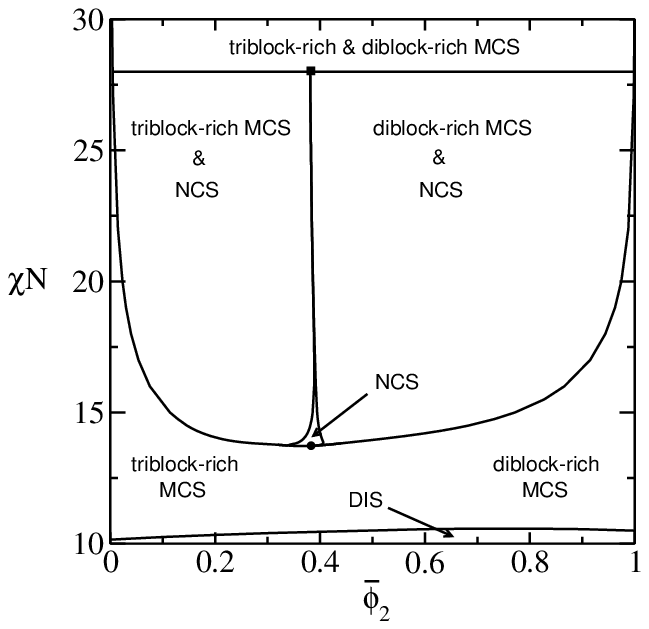}
\end{center}
{\small FIG.\ 2. The phase  diagram in  the $\chi N$---$\bar{\phi}_{2}$
plane, for the case where all the Flory-Huggins interaction parameters
are equal  ($\chi_{AB} =\chi_{AC}=\chi_{BC} \equiv  \chi$). The
triblock to diblock length ratio is $\Omega = 1.5$. The labels
are  as follows:  DIS:  disordered; MCS:  mixed centrosymmetric;  NCS:
non-centrosymmetric.  Below the phase  coexistence line, the MCS phase
goes  from being  triblock-rich to  diblock-rich,  as 
the average diblock volume fraction $\bar{\phi}_{2}$
increases.   The critical  point  for  the MCS  to  NCS transition  is
indicated by a solid circle.  The narrow region of NCS phase stability
terminates at a triple point,  indicated by a solid square.  
Regions of two-phase coexistence are indicated.
Above the
horizontal  line  at  $\chi  N  \approx  28$  there  is  two-phase
coexistence between an  almost pure triblock phase and  an almost pure
diblock phase.  }
\label{FIG:CHIEQ1}
\end{figure}

For larger values of $\chi N$ (in the intermediate segregation regime 
around $\chi N \approx 14$ ) 
the B block  tends to expel the diblock a and
c blocks from its domain,  and it becomes favourable for the triblocks
and  diblocks to  spatially separate.   The transition  from  mixed to
spatially-separated  states is  indicated  
by the phase coexistence line in Fig.\  2.
Above  the  transition,  the  existence  of  a  stable  NCS
phase of alternating triblock  and diblock layers (...ABCca... as in 
Fig.\ 1c) in a narrow  region  around
$\bar{\phi}_{2}  \approx 0.4$  prevents the system from phase separating
into coexisting triblock-rich and diblock-rich phases. Instead, for values
of $\bar{\phi}_{2}$ lower than the stability region for the NCS phase,
the triblock-rich phase coexists with  the NCS phase, while for values
of $\bar{\phi}_{2}$  higher than this  stability region the  NCS phase
coexists  with the  diblock-rich phase.  The volume  fraction  of each
coexisting  phase is  given by  the lever  rule.   As $\bar{\phi}_{2}$
increases and the longer triblocks are removed, the periodicity $D$ of
the  structure  decreases.  Thus,   for  a  given  $\chi  N$,  the
triblock-rich MCS  phase has a longer  period than the  NCS phase, and
the NCS phase  has a longer period than the  diblock-rich MCS phase. A
typical  density  profile  for  the   NCS  phase  is  shown  in  Fig.\
4.
\vspace{.25 in}
%
%
\begin{figure}
\begin{center}
    \leavevmode
    \epsfxsize=3.0 in
    \epsffile{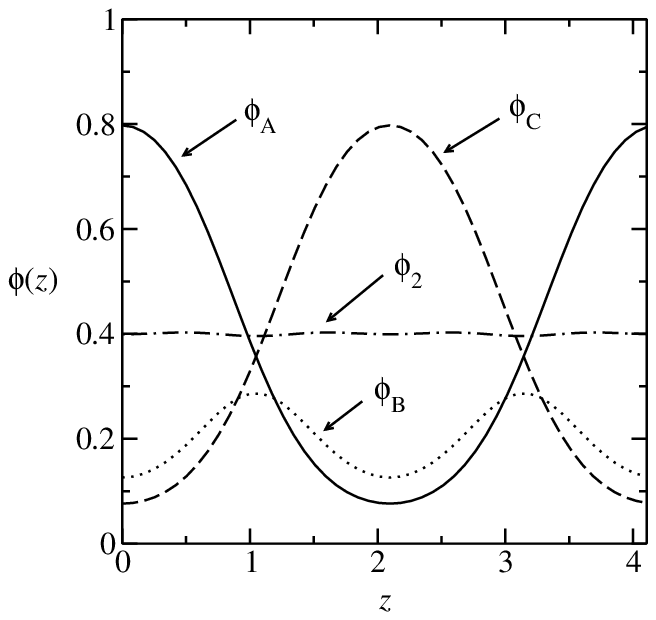}
\end{center}
{\small FIG.\ 3. Profile of one period of the MCS lamellar phase for $\chi N= 13$
(all Flory-Huggins interaction parameters are equal), 
$\bar{\phi}_{2} = 0.4$ and $\Omega = 1.5$. The distance perpendicular to the 
lamellae is $z$, measured in units of the radius of gyration of the 
diblock, $R_{g2}$.}
\label{FIG:MCS}
\end{figure}

The  widest region  of NCS  stability occurs  near the  bottom  of the
phase-coexistence curve. Figure  5 shows this region in
more detail.   The phase boundaries appear  to converge at  a point where
$\bar{\phi}_{2} \approx 0.365$ and  $\chi N \approx 13.72$.  Since
the free-energy  differences between the structures become very
small near this  point and the NCS phase is  only slightly stable with
respect to the MCS phase, it becomes difficult to numerically compute
the  phase boundaries  in  this  region.  The  dashed  lines in  Fig.
5  are  thus extrapolations  of  the phase  boundaries.
Given the fact  that the free-energies of the   structures appear
to merge at  this point, and the fact that  the difference between the
period of the  stable NCS and the metastable  MCS structure approaches
zero  as  this point  is  approached, we  believe  that  the point  of
convergence is a second-order critical point.  The order-parameter for
this transition is the amplitude of the odd-parity basis functions 
(sines) for the monomer profiles.
\vspace{.25 in}
\begin{figure}
\begin{center}
    \leavevmode
    \epsfysize=3.0 in
    \epsffile{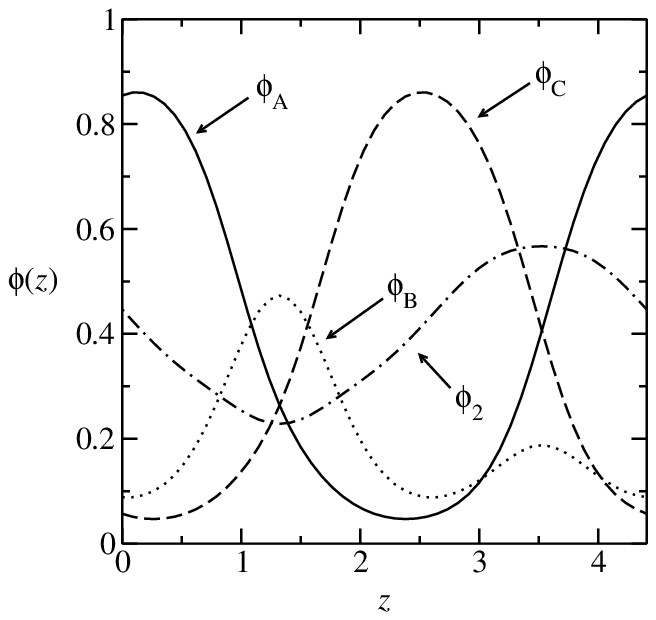}
\end{center}
{\small FIG.\ 4. Profile of one period of the NCS lamellar phase for $\chi N= 14$
(all Flory-Huggins interaction parameters are equal), 
$\bar{\phi}_{2} = 0.4$ and $\Omega = 1.5$. The
distance perpendicular to the 
lamellae is $z$, measured in units of the radius of gyration of the 
diblock, $R_{g2}$. }
\label{FIG:NCS}
\end{figure}
When $\chi N > 17$   
the  region of  NCS
stability narrows to a sliver, as shown in Fig.\ 2. The
NCS stability region terminates at a triple point at $\bar{\phi}_{2}
\approx 0.381$ and $\chi N \approx 28.0$ where the triblock-rich,
diblock-rich and  NCS phases coexist.  For larger values of $\chi N$ the
NCS phase is unstable, and there is only two-phase coexistence between
almost  pure  triblock and  diblock  phases.   This  
part of the phase diagram  is
similar to  the phase diagram of  a eutectic material,  where a liquid
phase (here  the NCS  phase) intervenes between  two solid  phases 
(here the MCS phases) and
terminates at a eutectic point (here the triple point) \cite{KITTEL80}.

We  have slightly  varied  the relative  values  of the  Flory-Huggins
interaction parameters and examined  the effects on the phase diagram.
To try to avoid non-lamellar  phases, we have kept the interactions in
the triblock symmetric ($\chi_{AB}=\chi_{BC}$) and examined the effect
of having only slightly weaker  repulsion between the middle and outer
blocks ($\chi_{AB}=\chi_{BC}=0.8\mbox{ } \chi_{AC}$) and only slightly
stronger  repulsion  between  these  blocks  ($\chi_{AB}=\chi_{BC}=1.1
\mbox{ } \chi_{AC}$).  The value $\Omega=1.5$ is used.
It is known, however, that for large variations
in  the Flory-Huggins interaction  parameters non-lamellar  phases may
arise  (such as  B  cylinders or  spheres  on AC interfaces  when
$\chi_{AB}=\chi_{BC}>\chi_{AC}$) \cite{ZHENG95}.
%
%
\begin{figure}
\begin{center}
    \leavevmode
    \epsfysize=3 in
    \epsffile{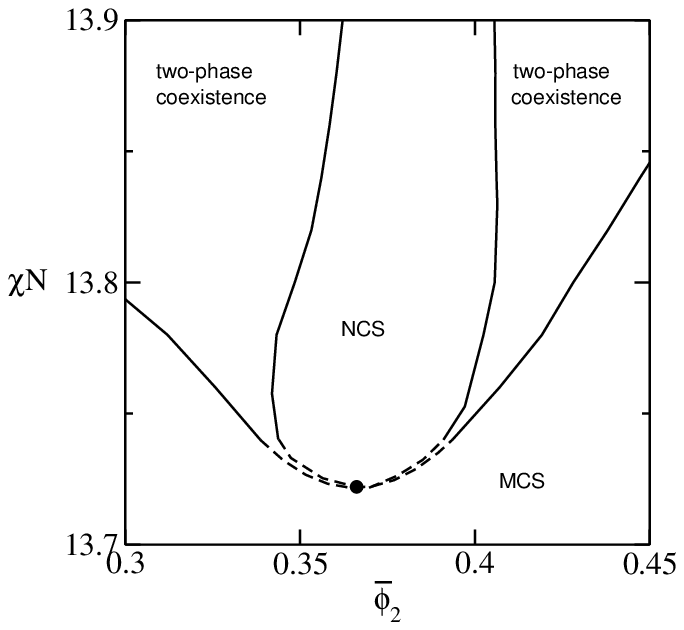}
\end{center}
{\small FIG.\ 5. The region in Fig.\ 2
near the critical point in more detail. 
The parameters and notation are the same as in 
Fig.\ 2. The dashed lines are extrapolations of the phase 
boundaries.
}
\label{FIG:CHIEQ2}
\end{figure}

In Fig.\ 6 the  phase diagram in the 
$\chi_{AC} N$---$\bar{\phi}_{2}$    plane   is   shown    for   the    case   
where
$\chi_{AB}=\chi_{BC}= 0.8  \mbox{ } \chi_{AC}$.  The  
overall structure of the
phase diagram 
is   the  same  as   in Fig.\ 2 where the  
Flory-Huggins  interaction parameters were equal
(to emphasize the region around the critical point, we only show the lower 
part of the phase diagram in Fig.\ 6).  The
region  of NCS  stability  is  shifted to  slightly  higher values  of
$\bar{\phi}_{2}$.   More  noticeably, the  phase  coexistence line  is
shifted to higher values of  $\chi_{AC} N$. The critical point now
occurs at  $\bar{\phi}_{2} \approx 0.414$  and $\chi_{AC}N \approx
16.79$ and  the triple  point occurs at  $\bar{\phi}_{2}\approx 0.391$
and $\chi_{AC}N  \approx 48$. Since the repulsion  between the B
block and the diblock a and c blocks drives the triblocks and diblocks
to  spatially separate,  decreasing  this repulsion  raises the  phase
coexistence line. In Fig.\ 7 the phase diagram in the
$\chi_{AC}  N$---$\bar{\phi}_{2}$ plane  is shown  for  the case
where $\chi_{AB}=\chi_{BC}= 1.1 \mbox{  } \chi_{AC}$. Again, the basic
structure   of   the   phase   diagram   is   unchanged   from   Fig.\
2, while the phase coexistence line is shifted to lower
values of $\chi_{AC}  N$.  The region of NCS  stability is shifted
to  slightly lower  values  of $\bar{\phi}_{2}$.   The critical  point
occurs  at $\bar{\phi}_{2} \approx  0.350$ and  $\chi_{AC}N \approx
12.69$ and  the triple  point occurs at  $\bar{\phi}_{2}\approx 0.376$
and  $\chi_{AC}N \approx 23$.   The increased  repulsion between
the B  block and the diblock a  and c blocks drives  the triblocks and
diblocks to  spatially separate at lower values  of $\chi_{AC} N$,
lowering the phase  coexistence line. Interestingly, the phase coexistence
line in Fig.\ 7 is more asymmetric than the phase 
coexistence lines in either Fig.\ 2 or Fig.\ 6.
This is due to the increased repulsion of the B block in 
Fig.\ 7, which accentuates the difference between the 
triblock and the diblock.
%
%
\begin{figure}
\begin{center}
    \leavevmode
    \epsfxsize=3.0 in
    \epsffile{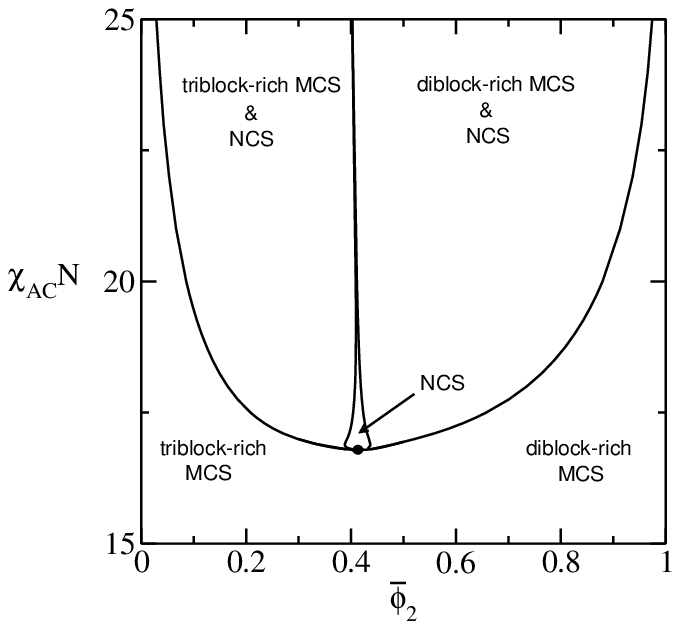}
\end{center}
{\small FIG.\ 6. The phase diagram in the $\chi_{AC} N$---$\bar{\phi}_{2}$ 
plane, for the case where $\chi_{AB}=\chi_{BC}=0.8 \mbox{ } \chi_{AC}$. The
triblock to diblock length ratio is $\Omega = 1.5$. The notation used is the
same as in Fig.\ 2. We have chosen to focus on the region of 
the phase diagram near the critical point.
The disordered phase exists below 
$\chi_{AC} N \approx 10.5$, and the triple point exists for 
$\chi_{AC} N \approx 48$.
}
\label{FIG:MIDLOUT}
\end{figure}

It was hoped  that by looking at variations in  the relative values of
the Flory-Huggins  interaction parameters the region  of NCS stability
could  be  expanded beyond  that  of  Fig.\  2. In  the
parameter  range we  have  examined,  we have  found  little, if  any,
dependence  of  the  width  of   the  NCS  stability  region  on  such
variations. It appears that, other  than an upward (or downward) shift
of the phase coexistence curve and triple point,  
the structure of the phase diagram is
insensitive  to  slight  variations  in  the relative  values  of  the
Flory-Huggins interaction parameters.

To examine  the effect of varying the  relative 
degrees of polymerization 
of the triblocks and diblocks we  show the phase diagram in the 
$\Omega$---$\bar{\phi}_{2}$ plane in Fig.\  8. In this case, the
Flory-Huggins parameters  are constant and  equal: $\chi N  = 15$.
Since an increase  in the triblock polymerization index  results in an
increased  effective segregation  in the  triblock, the  B  block will
expel  the diblock a  and c  blocks for  large enough  $\Omega$.  Thus
Figs.\ 2 and 8 are similar.   There is a
critical point  at $\bar{\phi}_{2} \approx 0.391$  and $\Omega \approx
1.34$ and a triple point at $\bar{\phi}_{2} \approx 0.370$ and $\Omega
\approx 1.72$).   It is significant  that the region of  NCS stability
slopes to  lower $\bar{\phi}_{2}$ as  $\Omega$ is increased,  and that
this region roughly corresponds to the condition that the number of
diblock copolymers in the blend equals the number of triblock 
copolymers,
%
%
\be
\bar{\phi}_{2}^{*} = \frac{1}{1 + \Omega}.
\label{EQ:EACOND}
\ee
Equation  (\ref{EQ:EACOND}) is  plotted in  Fig.\ 8 for
comparison. For the phase  diagram in Fig.\ 2, where the
region of  NCS stability  is almost vertical,  Eqn.\ (\ref{EQ:EACOND})
with  $\Omega=1.5$ gives $\bar{\phi}_{2}^{*}=0.4$,  which is  close to
what is seen.
%
%
\begin{figure}
\begin{center}
    \leavevmode
    \epsfxsize=3.0 in
    \epsffile{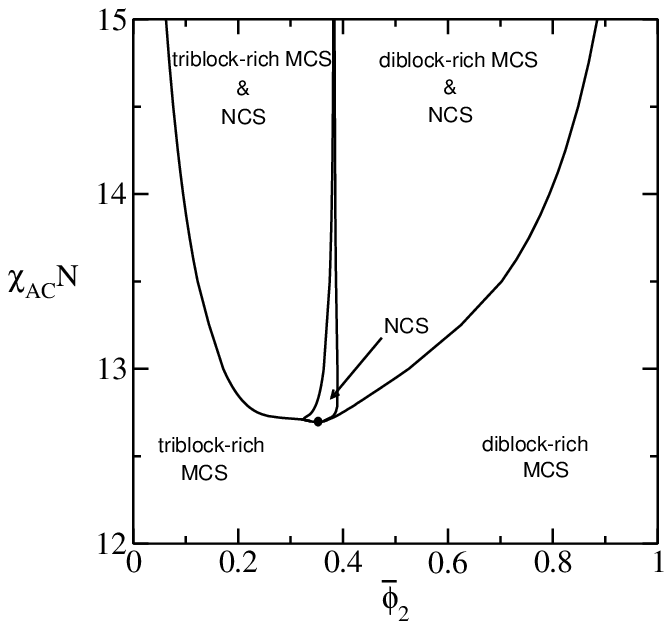}
\end{center}
{\small FIG.\ 7.
The phase diagram in the $\chi_{AC} N$---$\bar{\phi}_{2}$ 
plane, for the case where $\chi_{AB}=\chi_{BC}=1.1 \mbox{ } \chi_{AC}$.
The triblock to diblock length ratio is $\Omega = 1.5$.
The notation used is the
same as in Fig.\  2.
We have chosen to focus on the region of 
the phase diagram near the critical point.
The disordered phase exists below 
$\chi_{AC} N \approx 10.5$, and the triple point exists for 
$\chi_{AC} N \approx 23$.}
\label{FIG:MIDGOUT}
\end{figure}
%
%
%
\begin{figure}
\begin{center}
    \leavevmode
    \epsfxsize=2.9 in
    \epsffile{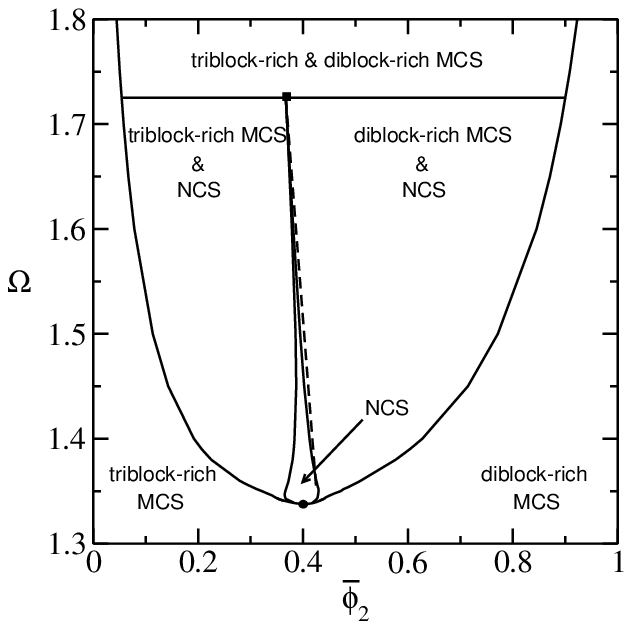}
\end{center}
{\small FIG.\ 8. The phase diagram in the $\Omega$---$\bar{\phi}_{2}$ plane, 
for the case where all the Flory-Huggins interaction parameters are 
equal ($\chi N = 15$). The notation used is the
same as in Fig.\  2.
 The dashed straight line corresponds to Eqn. (\ref{EQ:EACOND}).
}
\label{FIG:OMEGA}
\end{figure}
%
\pagebreak
We  now comment  on the  stability of  the centrosymmetric  (CS) phase
consisting   of  alternating   triblock   and  diblock   double-layers
(...ABCcaacCBA...).  Naively,  one might expect  that a flip  of every
other period of the NCS phase would lead to a periodic structure which
is degenerate in free-energy with the original NCS phase, and has exactly
twice  the  period of  the  original  phase.   However, in  the  phase
diagrams   shown  in   Figs.\   2,  6, 7  and 8 
the CS structure  is metastable \cite{EXCEPTION}.  
Below the triple point
the CS  structure has a  slightly higher free-energy density  than the
NCS structure. Above the triple  point the CS structure has a slightly
lower  free-energy  density  than  the  NCS structure,  but  is  still
metastable  with respect  to two-phase  coexistence. This  makes sense
since, above  the  triple   point,  a  metastable  NCS  structure  may
phase-separate by going though a  series of CS phases of longer period
and  lower  free-energy.  We  also  observe  that  the period  of  the
metastable CS  structure is only  approximately twice that of  the NCS
structure.  Subtle  differences between the chain  conformations in aA
interfaces (where one chain comes  from the diblock and the other from
the  triblock)  and those  in  AA  (and  aa) interfaces  are  probably
responsible  for breaking  the degeneracy  of  the NCS  and CS  phases
\cite{LEIBLER99,BIRSHTEIN99}.

In  the experiments  of Ref.\ \cite{GOLDACKER99},  performed  with symmetric
block compositions and  $\Omega=1.5$, two-phase coexistence between a
NCS phase and a  triblock-rich phase were observed for $\bar{\phi}_{2}
\approx 0.18$, while for $\bar{\phi}_{2} \approx 0.4$ a predominantly
 NCS phase was observed.   
These results are consistent with  the phase structure
seen  in Fig.\  2.  However,  when comparing  theory to
experiment  it  is  important to  note  that  it  is likely  that  the
experiments  were performed in  the strong-segregation  regime
rather than  in the  intermediate segregation  regime ($\chi N \approx 14$)
where  we  predict the  widest
stability  region for  the NCS  phase. Also,  the assumption  in Fig.\
2  of equal  Flory-Huggins interaction  parameters does
not hold for the experiments. On the other hand, we have seen that the
structure of the phase diagram  is insensitive to slight variations of
the  Flory-Huggins interaction  parameters, with  the exception  of an
upward or downward  shift in the widest region  of NCS stability and
the triple point.  Thus it is possible for the experiments to be 
performed at relatively strong segregation, but still be below the 
triple point in the phase diagram. Also, although we cannot exclude the 
possibility that the experiments observed the pure NCS phase, it 
is likely, given the narrow region of NCS stability, that the
experiments were in the two-phase coexistence region and observed a 
predominance of NCS phase due to the proximity of the NCS 
stability region.

One possible experimental strategy
for obtaining a pure NCS phase is to 
begin in the disordered phase with the blend composition tuned to the 
composition of the critical point (here $\bar{\phi}_{2} \approx 0.35-0.4$) and 
then perform a shallow temperature quench into the NCS stability region
\cite{RUSSELL}.
The quench should be as shallow as possible so as to minimize the 
influence of both kinetic effects and the intervening MCS region  on the 
formation of the NCS phase. As we have shown in Fig.\ 7, 
when $\chi_{AB}= \chi_{BC} > \chi_{AC}$
the region of MCS 
stability narrows as the MCS-to-NCS critical point moves to 
lower values of $\chi_{AC} N$, closer to the order-disorder transition. 
However, our preliminary investigation into the possibility of a direct 
transition from the disordered to NCS phase suggests that for a large relative
repulsion between the triblock middle and outer blocks ($\chi_{AB}=\chi_{BC}
=1.5 \mbox{ } \chi_{AC}$) 
the  NCS phase can become unstable to phase separation
into triblock- and diblock-rich phases, while a narrow 
region of MCS stability remains. Thus a compromise must be found between 
decreasing the width of the MCS region and maintaining the
stability of the NCS phase.

%
%
\section{Conclusions}

We have examined the phase behaviour  of blends of ABC triblock and ac
diblock   copolymers    using   self-consistent   field    theory.   A
representative phase  diagram is  shown in Fig.\ 2. 
For segregations just  above  the order-disorder transition
the blend  forms a centrosymmetric lamellar phase  where the triblocks
and diblocks  mix together (the MCS phase).   For stronger segregation
(increased $\chi  N$ or increased  $\Omega$) the B block  tends to
expel the  diblock a and  c blocks from  its domain. As a  result, the
triblocks   and  diblocks   spatially  separate   by   forming  either
alternating  layers  of  triblock  and diblock  or  a  phase-separated
state.  In  the  former  case,  we  observe  a  narrow  region  around
$\bar{\phi}_{2}  \approx   0.4$  where  a  pure   NCS  lamellar  phase
(Fig.  1c)  is  stable.   In the  latter  case,  we
observe large  regions of two-phase coexistence between  the NCS phase
and  either  a  triblock-rich  or  a diblock-rich  MCS    phase,
depending  on the value  of $\bar{\phi}_{2}$.   The structure  of this
phase  diagram is  insensitive to  slight variations  in  the relative
values of the Flory-Huggins interaction parameters. The only effect is
an upward  (or downward)  shift of the  phase coexistence  curve
and the triple point. Such
variations do not measurably effect the width of the NCS region.
 
For intermediate segregation, 
our results suggest that there is a second-order
critical  point  where a  continuous  transition  from  the MCS  phase
directly  to  the  NCS  phase  occurs. The order-parameter for
this transition is the amplitude of the odd-parity basis functions 
(sines) for the monomer profiles. As the segregation is increased,  
the region of  stability for the  NCS phase
narrows  to a  sliver, and  terminates at  a triple  point.  Above the
triple  point  there  is  two-phase coexistence  between  almost  pure
triblock and  diblock phases.  This is the  first work to  suggest the
existence  of the critical  point and  the triple  point in  the phase
diagram for this blend.

The  phase diagram in  Fig.\ 2  is consistent  with the
experiments  of  Goldacker   {\em  et  al.}   \cite{GOLDACKER99},  who
observed coexisting triblock-rich and NCS lamellae for $\bar{\phi}_{2}
\approx 0.18$, and a  predominance of NCS lamellae for $\bar{\phi}_{2}
\approx 0.4$. The  narrowness of the region of  NCS stability suggests
that  careful  adjustment of  the  blend  parameters  is necessary  to
observe the pure NCS phase. Our  theory predicts that  the width of  the NCS
stability   region  can   be  maximized   by studying blends near the 
phase coexistence curve which have  
 a  triblock to diblock length-ratio less
than 1.5.

Earlier theoretical  work by Leibler {\em et  al.\ } \cite{LEIBLER99} 
and Birshtein {\em et al.} \cite{BIRSHTEIN99} on
these blends  focussed on the  mechanism responsible for  the stability of
the  NCS phase in the  strong-segregation limit.   In Ref.\ \cite{LEIBLER99},
subtle   entropic    interactions   arising   from    the   asymmetric
interpenetration  of a  and A  blocks from  the diblock  and triblock,
respectively, cause the  system to favour mixed aA  interfaces over AA
(or  aa)  interfaces,  and lead  to  a  stable  NCS phase.  In Ref.\
 \cite{BIRSHTEIN99} the different grafting densities of the brush coming 
from the triblock and the brush coming from the diblock favoured 
mixed aA  interfaces.
Our  work
complements this  earlier work  since we can  examine the  blend phase
diagram without the need to assume the strong-segregation limit.  However, in
contrast to Refs.\ \cite{LEIBLER99} and \cite{BIRSHTEIN99}, our calculations 
show that the NCS phase
is   only  stable   for   intermediate  segregation,   and  that   for
strongly-segregated  blends there is  two-phase coexistence  between 
triblock-rich  and diblock-rich phases.  A reconciliation between our
results and those of Refs.\ \cite{LEIBLER99} and \cite{BIRSHTEIN99}
in the strong-segregation regime is  necessary.  
A  clue   to  the  nature  of  the  entropic
interaction driving the stability of  the NCS phase is our observation
that the region of NCS stability roughly corresponds to Eqn.\  
(\ref{EQ:EACOND}), the condition 
for there to be  equal numbers of triblock and 
diblock copolymers in  the two polymer brushes composing the aA interface.  It
would be interesting to examine  the interpenetration of the triblock and
diblock chain profiles at the aA interface using self-consistent field
theory.  This would allow a more direct comparison to be made with the
theories of Refs.\   \cite{LEIBLER99} and \cite{BIRSHTEIN99}. 
It may be that with increasing segregation the chain 
interpenetration at the aA interfaces becomes more symmetric and the 
brush grafting densities more similiar, 
reducing the stability of the NCS phase in the 
strong segregation regime and leading to the triple point seen in Fig.\ 2.
An  analytical calculation  of the
stability of  the MCS phase towards  the NCS phase,  near the critical
point, may also be illuminating.

Only lamellar structures are considered in this study.  Even though we
have been careful to select parameters that favour the lamellar phase,
it is possible that stable  non-lamellar structures exist in our phase
diagram (some observations of  non-lamellar structures in these blends
are discussed in  Ref.\ \cite{ABETZ00}).  We can however  say that, for
the parameters  examined, the size of  the region of  NCS stability is
the  {\em  maximum}  possible.   Also,  we  have  ignored  composition
fluctuations  in the  present mean-field  treatment.  Fluctuations may
modify some of the detailed structure of the phase diagram, especially
near the  critical point, but we  believe a region  (perhaps a smaller
one)  of  NCS  stability  will  survive  the  effects  of  composition
fluctuations.
%
%
\acknowledgements
The authors would like to thank Prof.\ Hugh Couchman for access to his 
computing resources and Mr.\ David Cooke for helpful suggestions with coding.
They would also like to thank Profs.\ Michael Schick,
David Morse and Thomas P.\ Russell for useful comments. This 
work was supported by the Natural Sciences and Engineering Research Council
of Canada.
%
%

\end{multicols}

\begin{references}
\bibitem{EXCEPTION1} NCS phases have also been observed 
in  blends of two triblock copolymers (see Ref. \cite{ABETZ00}).
\bibitem{EXCEPTION} We have performed a preliminary survey of the 
phase diagram for the case where all the Flory-Huggins parameters 
are different ($\chi_{AB} = 1.1 \mbox{ } 
\chi_{AC}$, $\chi_{BC} = 0.9 \mbox{ } \chi_{AC}$). 
Our results suggest that, in this case, the CS phase can become 
stable (and the NCS phase metastable)  with increasing segregation, 
but before the triple point is reached. However, in this phase diagram 
the possibility for non-lamellar phases complicates the interpretation.
\end{references}
\end{document}